\documentclass{emulateapj}

\slugcomment{ }

\shorttitle{Sub-Pixel Response Measurement of NIR Sensors}
\shortauthors{N. Barron \textit{et al.}}

\include{natbib}
\begin{document}


\title{Sub-Pixel Response Measurement of Near-Infrared Sensors}


\author{
N.~Barron\altaffilmark{1},
M.~Borysow\altaffilmark{1,2},
K.~Beyerlein\altaffilmark{1,3},
M.~Brown\altaffilmark{1},
C.~Weaverdyck\altaffilmark{1},
W.~Lorenzon\altaffilmark{1},
M.~Schubnell\altaffilmark{1},
G.~Tarl\'e\altaffilmark{1},
A.~Tomasch\altaffilmark{1} }


\altaffiltext{1}{Department of Physics, University of Michigan,
Ann Arbor, MI 48109}
\altaffiltext{2}{Now at Department of Physics,
University of Texas, Austin, TX 78712.}
\altaffiltext{3}{Now at Material Science Department, Georgia Tech,
Atlanta, GA, 30332}


\begin{abstract}
Wide-field survey instruments are used to efficiently observe large regions of
the sky.  To achieve the necessary field of view, and to provide a higher
signal-to-noise ratio for faint sources, many modern instruments are
undersampled. However, precision photometry with undersampled imagers requires a
detailed understanding of the sensitivity variations on a scale much smaller than
a pixel. To address this, a near-infrared spot projection system has been
developed to precisely characterize near-infrared focal plane arrays and to study
the effect of sub-pixel non uniformity on precision photometry. Measurements of
large format near-infrared detectors demonstrate the power of this system for
understanding sub-pixel response.
\end{abstract}


%

\section{Introduction}
The \emph{Hubble Space Telescope} (HST) has provided overwhelming evidence for
the power of a space-based platform for optical and near-infrared (NIR)
astronomical observations \citep{HST1}. In particular, access to diffraction
limited imaging, stable observing conditions, and low background levels have
revolutionized our understanding of the faint and distant universe.

Space-science instruments are evolving from the initial HST instruments that
provided narrow-field observations to large scale survey instruments capable of
providing HST quality (or better) data across large regions of the sky. The HST
began this trend toward large-scale surveys with the Advanced Camera for Surveys
(ACS) in 2002 \citep{ACS}. The ACS has enabled a series of relatively wide-field
surveys, such as the GOODS, GEMS, and COSMOS surveys, however even the largest of
these reaches only a few square degrees. These surveys provide deep,
high-resolution imaging, but a limited picture of the statistical properties of
the objects they detect, and little information about the large scale
distribution of these objects. For comparison, large ground based surveys, such
as the SDSS (10,000 square degrees) provide precise statistical information with
lower resolution.~\citep{SDSS}

Future space-science instruments will combine high-resolution with wide-field
imaging to provide new data sets for cosmology and astrophysics. Wide-field
imaging missions have been proposed to study such diverse topics as planets
(Kepler) \citep{Kepler}, microlensing (GEST) \citep{GEST}, and dark energy (e.g.
SNAP \citep{SNAP} and DESTINY \citep{Destiny}). The central science goals for
most of these missions depend on the ability to make precise photometric
measurements in imaging mode. For example, Kepler will monitor the brightness of
about $10^5$ stars, watching for the very slight dimming ($<1\%$) associated with
planetary transits. SNAP will achieve relative distance measurements with a
smaller than $2\%$  uncertainty by comparing the restframe B and V magnitudes of
type Ia supernovae.

To achieve wide-field imaging and high resolution at an affordable cost, many
modern instruments operate in a critically sampled, or undersampled mode. Such
observations require a detailed understanding of the detector response. Large
scale inter-pixel variations in detector response are characterized by a variety
of well established flat-fielding methods. However, no such methods exist for
small scale intra-pixel sensitivity variations, which introduce uncertainties in
the conversion between the detected signal and incident light.

\section{Instrument}
\label{instrument_section}

To study  the intra-pixel response in NIR sensors, an automated spot projection
system has been developed. This system, the ``Spot-o-Matic,'' is part of the
University of Michigan Near Infrared Detector Testing Facility. It is designed to
measure one-dimensional and two-dimensional sub-pixel response profiles in large
format NIR focal plane arrays (FPAs) which are used in wide-field surveys.
Typical pixel sizes\footnote{Pixel sizes as small as $9 \times 9\,\mu$m$^2$ are
now under development.} for such detectors are approximately $20 \times
20\,\mu$m$^2$. The Spot-o-Matic performs sub-pixel response measurements by
step-scanning a stable, micron-sized spot across a small region of the detector,
and recording the detector's response at each spot position. A computer
controlled $x$-$y$-$z$ stage allows for large high-resolution step-scans of $25$
to $50$ pixels with sub-micron motion control. As intra-pixel sensitivity
variations were expected to be at the few percent level, the system was designed
to achieve a relative accuracy of better than $1\%$. The Spot-o-Matic has
achieved this accuracy goal in  measuring the pixel response function (PRF) for
the commercially-produced large format NIR sensors listed in
Table~\ref{Table:device-list}.

\subsection{Overview of Technique}

\begin{figure*} [htbp]
\centerline{\includegraphics[width=0.98\linewidth]{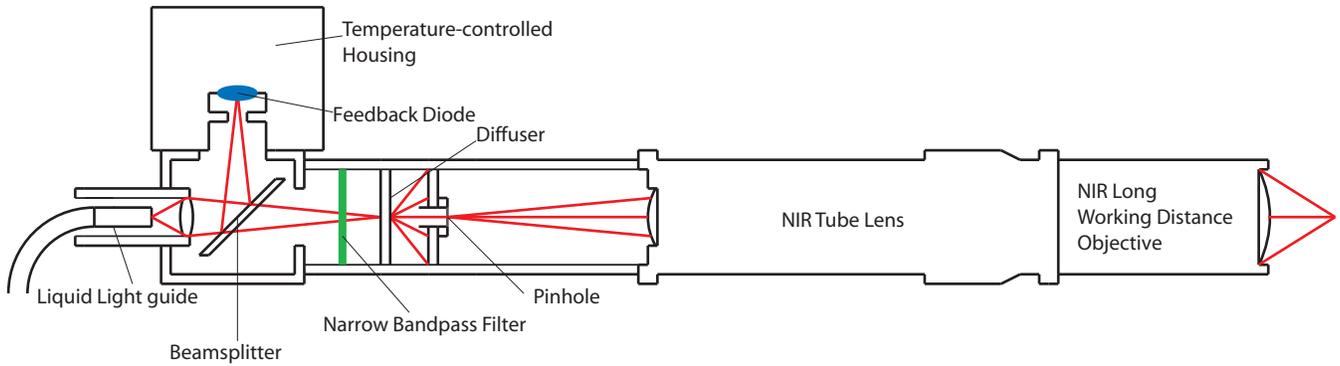}}
\caption[]{Schematic diagram of the optical components of the Spot-o-Matic.}
\label{Fig:Spot-o-Matic-optics} \vspace{0.05cm}
\end{figure*}

The design for the Spot-o-Matic is based on the concept of a pinhole projector
for visible light that was developed to measure diffusion in Charge Coupled
Devices (CCDs)  \citep{LBL-CCD}. This concept has been extended to an
instrument that operates both at visible and NIR wavelengths. The principle
optical components of the instrument are outlined in
Fig.~\ref{Fig:Spot-o-Matic-optics}.

NIR detectors designed for low background applications must be cooled to limit
the generation of thermal carriers (dark current) to acceptable values.
Therefore, the Spot-o-Matic utilizes a long working distance objective to
project a spot onto a NIR detector which is mounted inside a cryogenic dewar
(not shown in the figure). The Spot-o-Matic light source is a 50W QTH
(quartz-tungsten-halogen) lamp (Ushio BRL 12V50W) powered by a Spectra Physics
power supply (Model No.~69931) and operated in constant current mode. This mode
of operation provides better than $1\%$ stability and ensures that light
intensity variations are accurately decoupled from any detector response
variations.

The lamp is enclosed in a Photomax housing (Spectra Physics Model No.~60100)
with an elliptical reflector that focuses the light into a liquid light
guide.\footnote{Traditional fiber optics absorb light in the NIR, but liquid
light guides have high transmission through both the optical and NIR spectrum.}
This guide directs the light into an optical assembly, where it is divided by a
$30/70$ beam splitter (see Fig.~\ref{Fig:Spot-o-Matic-optics}). The weaker beam
is directed to a silicon feedback diode, while the stronger beam passes first
through a narrow bandpass filter and then through a diffuser before it
illuminates the pinhole. Light passing through the pinhole travels through a
NIR tube lens before it is focused by a long working distance NIR microscope
objective, through the dewar window, onto a cold sensor (not shown in
Fig.~\ref{Fig:Spot-o-Matic-optics}). The demagnification of the pinhole image
is set by the distance of the pinhole from the microscope objective. The
results presented here use a $10\,\mu$m pinhole and a $20\times$
demagnification. With this configuration, the spot point spread function (PSF)
is nearly diffraction limited at wavelengths  above $1000\,$nm.

The focusing microscope objective is a Mitutoyo infinity-corrected long working
distance objective,\footnote{The working distance is $31\,$mm, with a numerical
aperture (NA) of 0.26} chromatically corrected from $480\,$nm to $1800\,$nm for
both NIR and visible operation. Spot sizes as small as $\sigma=1\,\mu$m have
been achieved at a wavelength of $1050\,$nm. Spot sizes above the diffraction
limit can be produced by varying the pinhole aperture and the demagnification.
A set of narrow-bandpass filters ($10\,$nm FWHM) allows for stable
near-monochromatic sampling of the spectral sensitivity range of the tested
detectors.

The Spot-o-Matic optics are mounted on an NAI (National Aperture Inc.) Extended
Motorized MicroMini Stage (Model No.~MM-4M-EX-80) which supports
high-precision, high-resolution $x$-$y$-$z$ motion control.\footnote{The $z$
axis is the focus axis, $x$ and $y$ scan the projected spot across detector the
detector's columns and rows.} The $x$ and $y$ axes have a minimum step size of
$0.075\,\mu$m, with a positioning error of less than $1\,\mu$m per $2.54\;$cm
of linear travel, and a repeatability of $\pm 0.5\,\mu$m. To ensure a
reproducible focus, the $z$ axis uses a high-resolution linear encoder with a
step size of $20\,$nm. This provides an accurate reading of the stage position,
and avoids gear-head backlash when changing the direction of travel. The stage
is controlled by an NAI Motion MC-4SA Servo Amplifier System. Custom LabView
interfaces automate the test procedures.

The Spot-o-Matic assembly is mounted on an optical table inside a dark
enclosure to reduce the background photon flux. The tested sensors have a
cutoff wavelength between $1550\,$nm and $1750\,$nm and are thus sensitive to a
large background (a few thousand photons/pixel/second) of thermal radiation
from the optics and the walls of the dark enclosure at room temperature
($297\,$K). To reduce the thermal photon background, an effort was made to
place a cold narrow-band filter inside the dewar; however, due to space
constraints a filter could not be accommodated. Thus the intensity of the
projected spot was raised to a level at which the measurement was no longer
dominated by thermal radiation. For typical spot intensities and room
temperature background fluxes, the statistical uncertainty limit due to shot
noise on the source plus background photons is less than $1\%$.

\subsection{Beam Spot Characterization}

In order to measure intra-pixel responses, the Spot-o-Matic is designed to
produce spots much smaller than typical NIR pixels. For the measurements reported
here, scans were performed at wavelengths of $1050\,$nm and $1550\,$nm. These two
wavelengths were chosen to probe both the short and long wavelength response of
the detectors. $1550\,$nm was chosen because the long wavelength cutoff of the
InGaAs detector, which is $1.7\,\mu$m at room temperature, drops to $1.6\,\mu$m
at $140\,$K due to an increased bandgap energy at lower temperatures. $1050\,$nm
is near the approximately $850\,$nm cutoff of those detectors that do not have
the substrate removed.

\begin{figure}[bhtp]
\begin{center}
\includegraphics[width=0.95\linewidth]{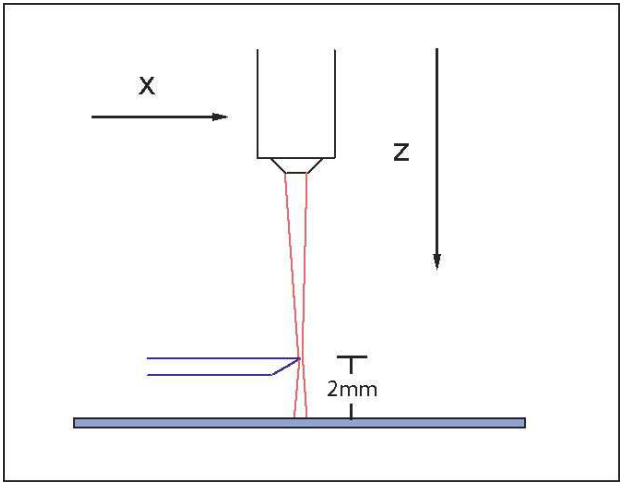}
\caption{Schematic diagram of the Foucault knife-edge scanning
  procedure. The point source image is scanned across a precision edge
  in the $x$ direction to determine the line spread function. The
  image location is found by focusing the beam in the $z$ direction to
  minimize the spot size. The unobscured beam typically extends over a
  few hundred pixels}
\label{Fig:knife-edge}
\end{center}
\end{figure}

The beam profile is measured using the Foucault knife-edge technique, a procedure
commonly used to determine the spatial profiles of images from point sources
\citep{knife-edge}. The transmitted beam intensity is recorded as a spot is
stepped across a razor blade mounted $2\,$mm above the detector surface, as shown
in Fig.~\ref{Fig:knife-edge}. Since the unobscured beam covers a few hundred
pixels, sensitivity variations within pixels do not affect the measurement.

\begin{figure*}[htbp]
\begin{center}
\includegraphics[height=0.45\linewidth, angle=270]{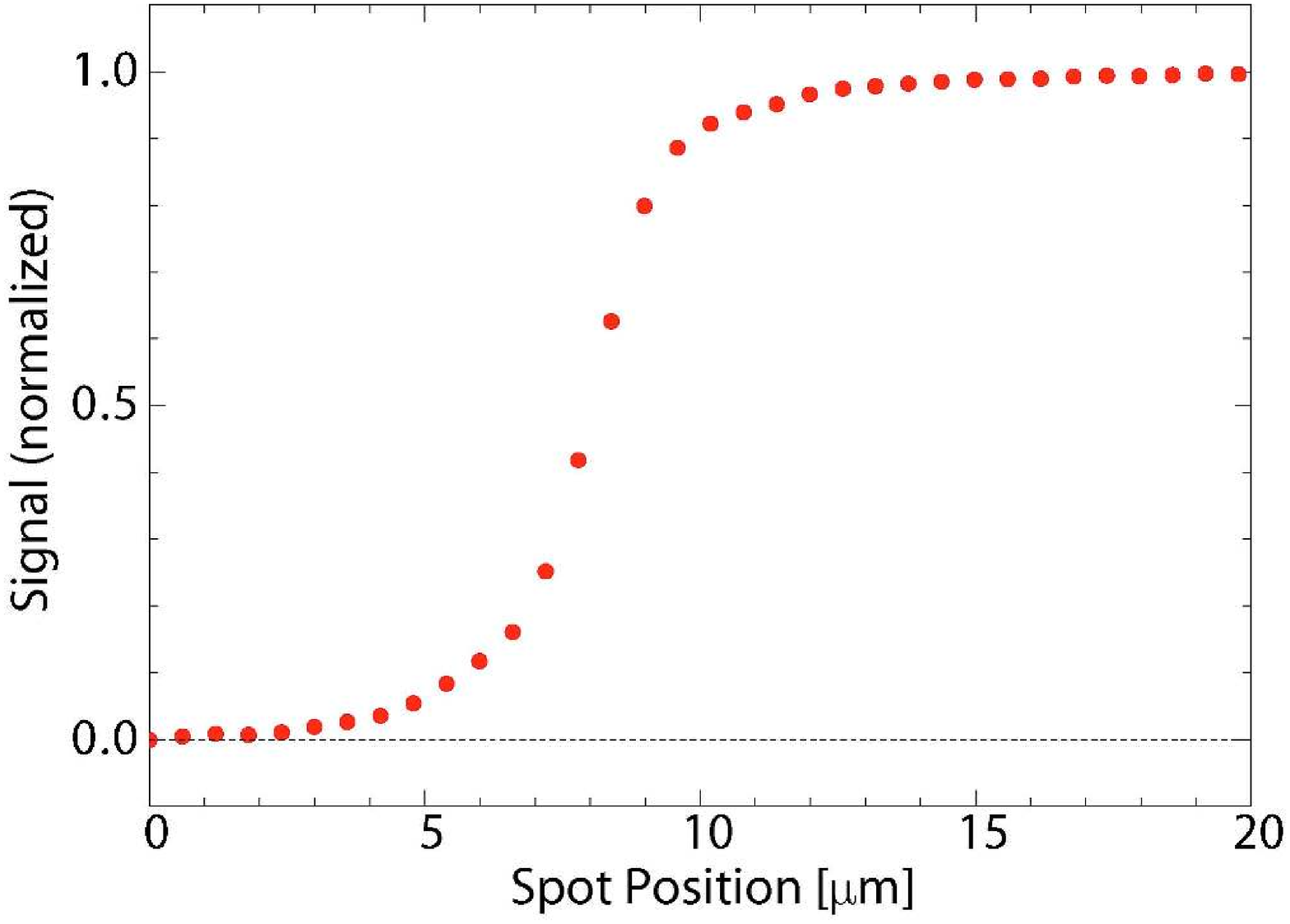}
\ \hspace*{12mm} \
\includegraphics[height=0.45\linewidth, angle=270]{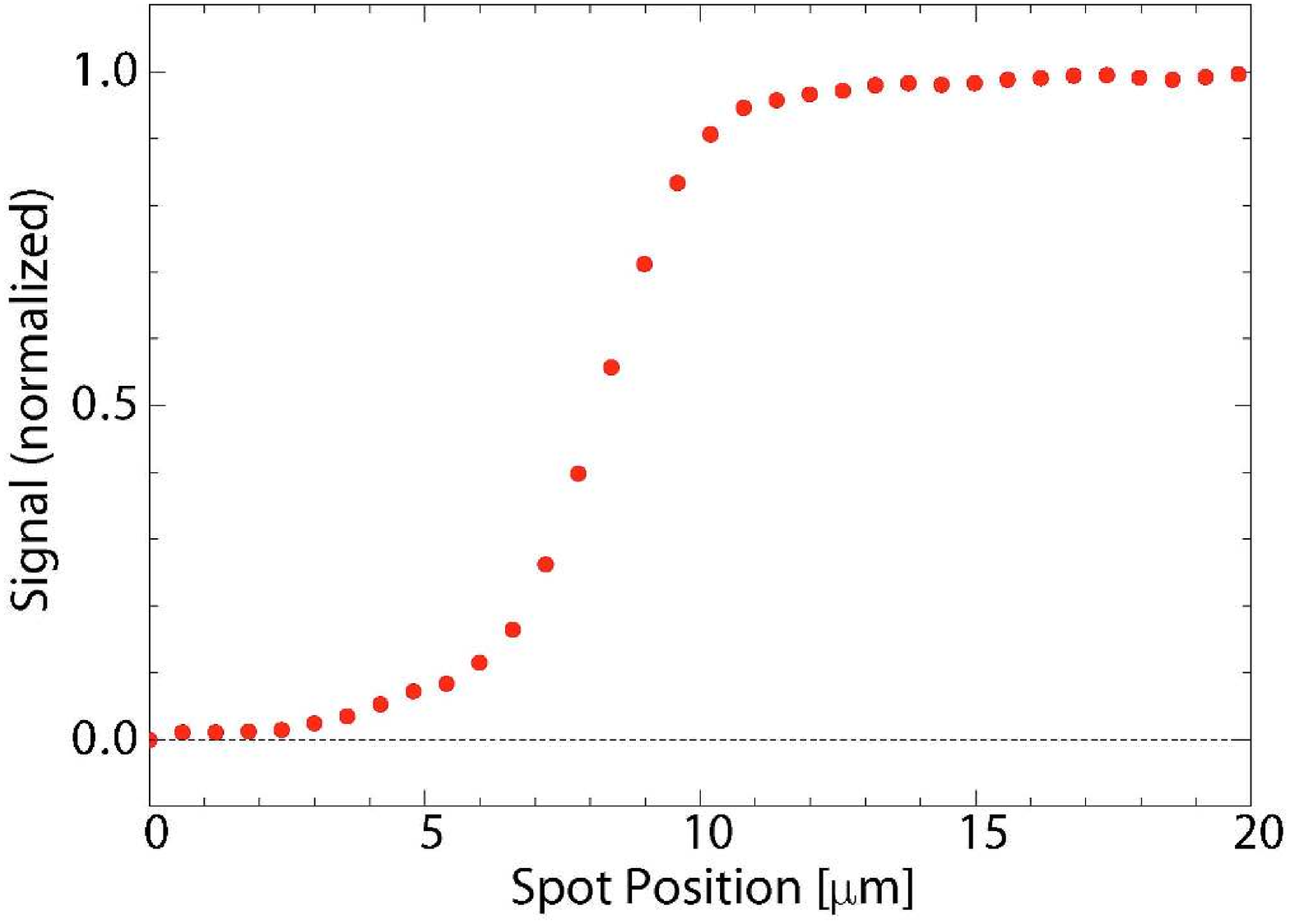}\\[5mm]
\includegraphics[height=0.45\linewidth, angle=270]{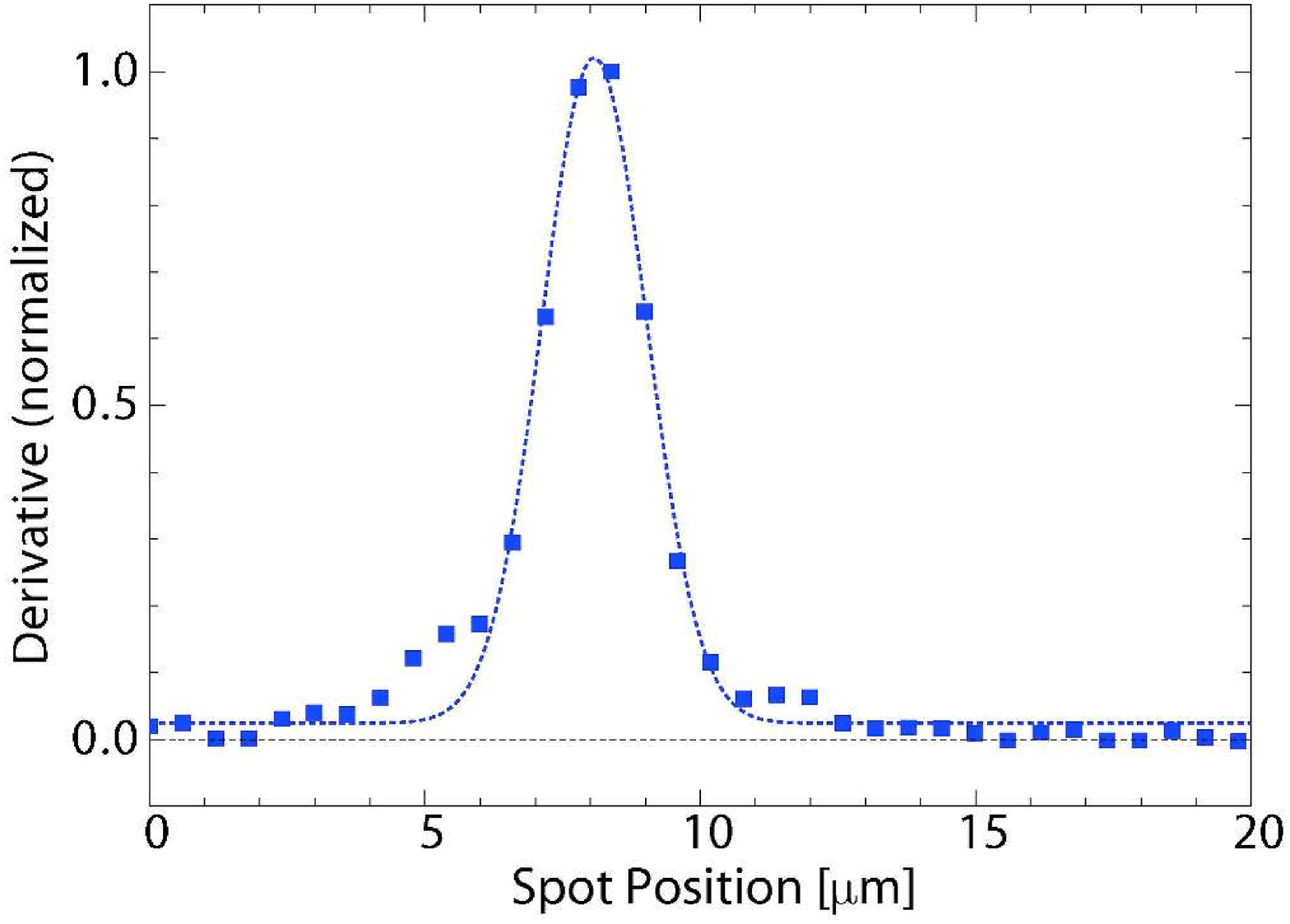}
\ \hspace*{12mm} \
\includegraphics[height=0.45\linewidth, angle=270]{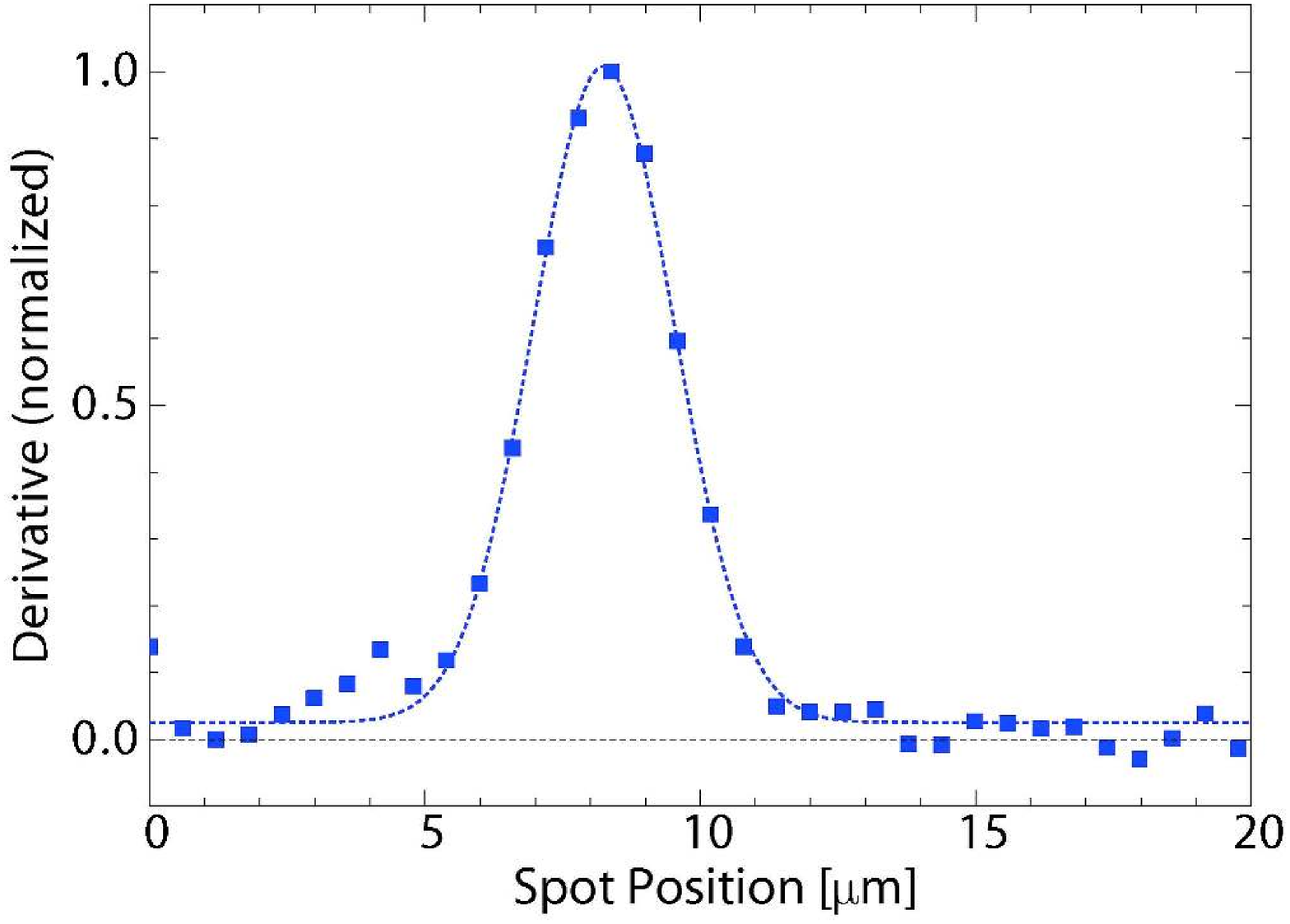}
\caption{Knife-edge scans at a wavelength of $1050\,$nm (left panel)
  and at $1550\,$nm (right panel).  The two top panels show the
  measured beam intensity as a function of the spot position.  The two
  bottom panels show the corresponding derivatives. Fitting a Gaussian
  function to the bottom plots (dotted curves) yields a spot size of $\sigma = 0.95
  \pm 0.03\,\mu$m for the $1050\,$nm wavelength and  $\sigma=1.28 \pm
  0.04\,\mu$m for the $1550\,$nm wavelength.}
\label{Fig:knife-edge-scans}
\end{center}
\end{figure*}

Figure \ref{Fig:knife-edge-scans} shows the results from a series of knife-edge
scans performed at wavelengths of $1050\,$nm and $1550\,$nm. The top panels in
this figure show the transmitted light intensities, or edge traces, as a function
of the beam's horizontal position. The bottom panels in
Fig.~\ref{Fig:knife-edge-scans} show the spatial profiles, or line spread
functions (LSFs), obtained by differentiating the edge traces. Note the small
``shoulder'' on the left side of each LSF. This shoulder likely results from
stray light reflected off the razor blade.

A Gaussian function\footnote{A correct description of the spot profile is the
one-dimensional integral of a two-dimensional Airy disk convolved with the
demagnified geometric pinhole image; however, the Gaussian function provides a
sufficiently good approximation.} provides an excellent fit to the central spot
profile, as shown in Fig.~\ref{Fig:knife-edge-scans} (the bottom panels). At a
wavelength of $1050\,$nm, the LSF has a fitted width of $\sigma = 0.95 \pm
0.03\,\mu$m, while at $1550\,$nm the width has increased to $\sigma=1.28 \pm
0.04\,\mu$m.

The expected widths of each LSF can be calculated by convolving the Airy disk
with the demagnified geometric pinhole image. The full width at half maximum
(FWHM) of the Airy disk is given by
\begin{equation}
\rm FWHM = 1.03 \,\lambda \, \frac{\sqrt{1 - NA^2}}{2\, NA},
\end{equation}
where the numerical aperture, NA, is 0.26 for the Spot-o-Matic optics. The root
mean squared (rms) width of a circular spot is $\sigma = 0.82\, d$, where $d =
0.5\,\mu$m is the diameter of the demagnified pinhole image in the absence of
diffraction. Adding the two components in quadrature yields an expected spot
size of $0.94\,\mu$m at a wavelength of $1050\,$nm and $1.32\,\mu$m at
$1550\,$nm. These predictions are in excellent agreement with the measured
values.

\subsection{Pixel Response}

To focus the spot onto the detector surface, a virtual knife-edge procedure is
employed. This method is analogous to the Foucault knife-edge procedure, but
does not use a razor blade to obstruct the beam. Instead, a sub-pixel size spot
is step-scanned across the center of an individual pixel while the signal in
that pixel is recorded. The edge of the pixel serves as the edge of the razor
blade in the Foucault knife-edge scan, and best focus is achieved at minimum
edge width. Although diffusion between pixels widens the edge, it remains
constant and is independent of the spot size.

\begin{figure}[htbp]
\begin{center}
\includegraphics[height=0.95\linewidth, angle=270]{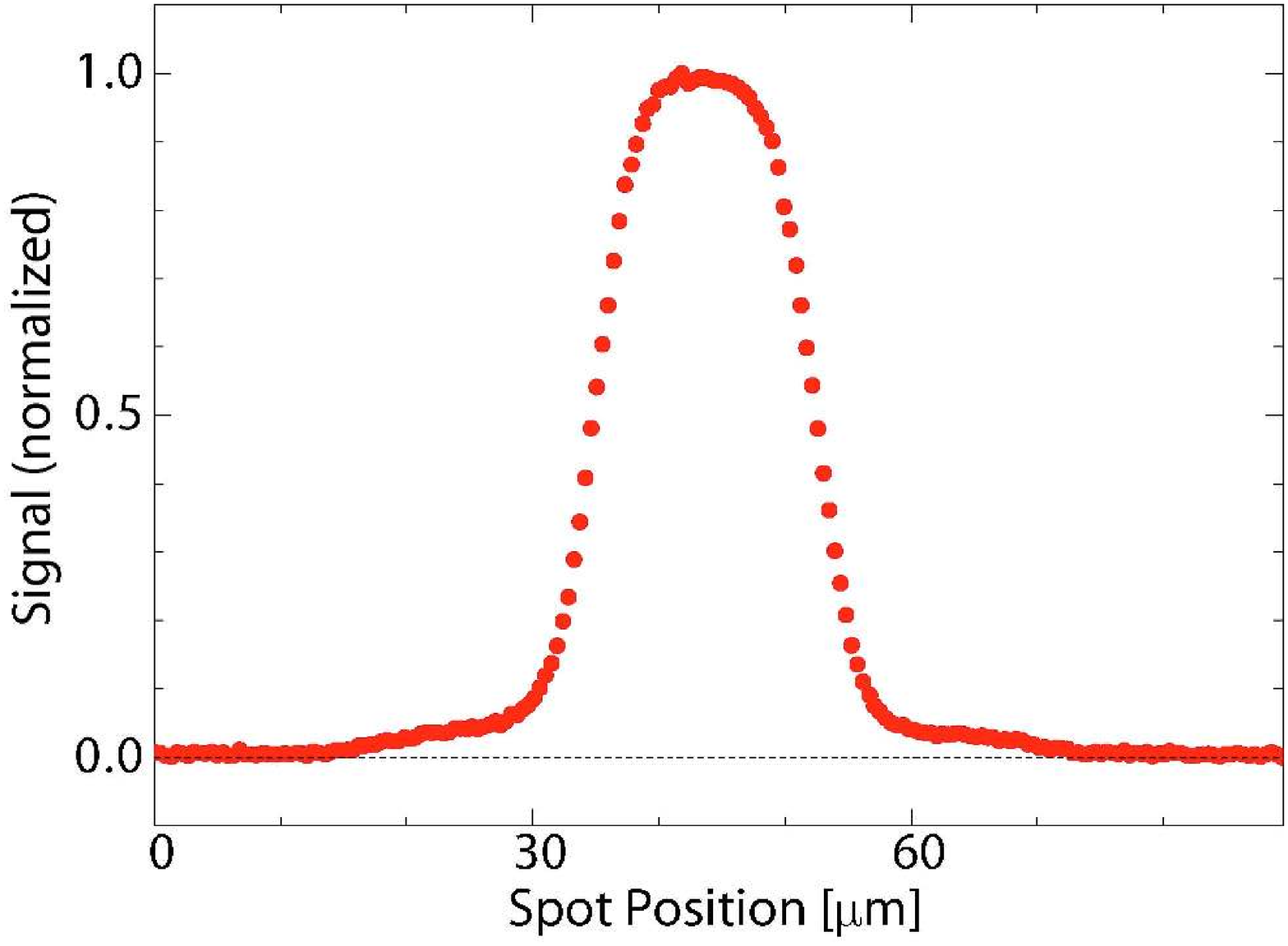}\\[5mm]
\includegraphics[height=0.95\linewidth, angle=270]{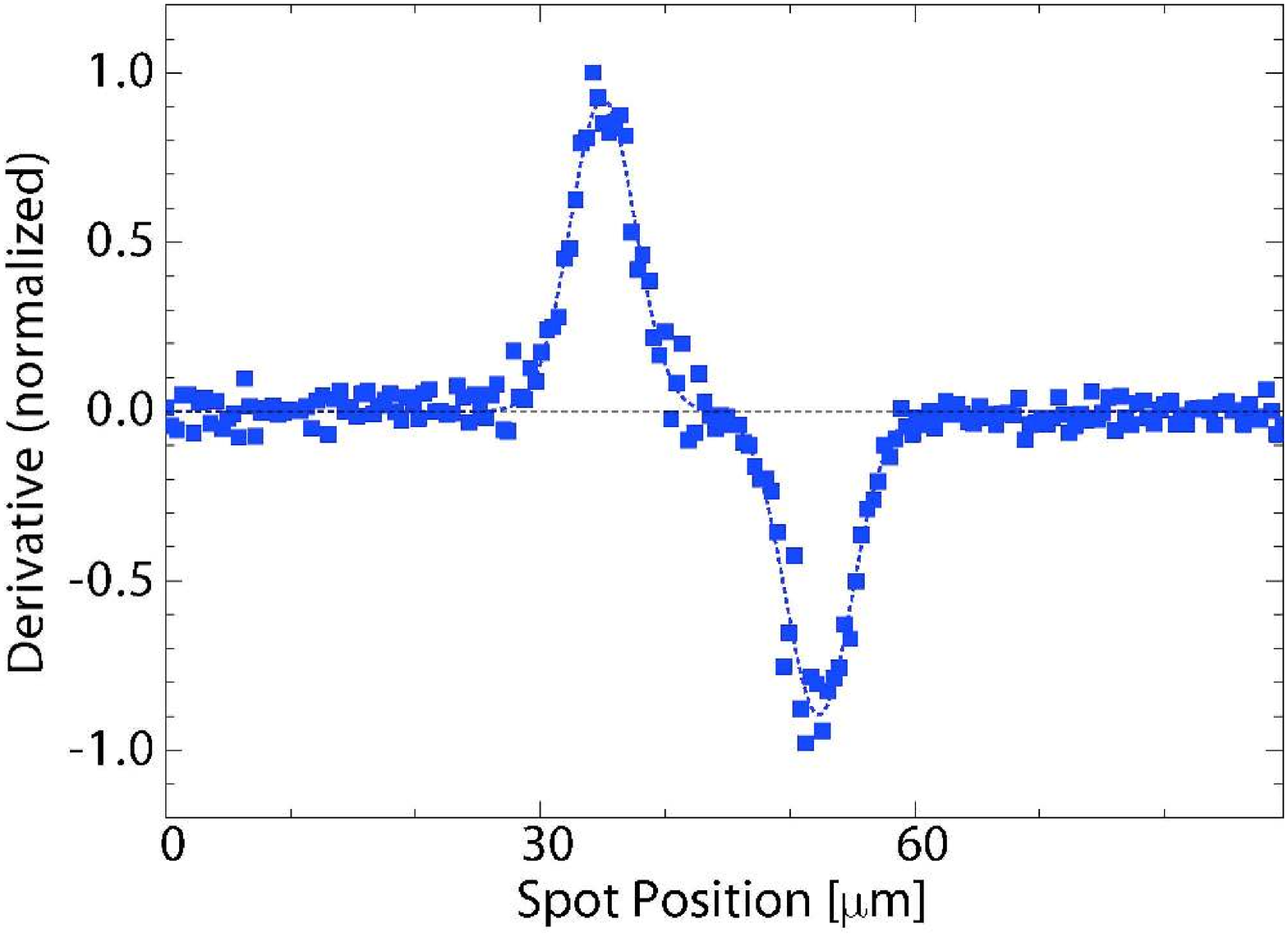}
\caption{Top panel: One-dimensional scan of an arbitrary single pixel
  for $1050\,$nm light. Bottom panel: The derivative of the signal
  with best-fit Gaussian functions over-plotted (dotted curve). The average width
  ($\sigma$) of the two Gaussians is $2.6\,\mu$m.}
\label{Fig:1d-virtual-knife-edge-scans}
\end{center}
\end{figure}

Figure \ref{Fig:1d-virtual-knife-edge-scans} shows the intensity profile for a
single representative pixel and its derivative at a wavelength of $1050\,$nm. The
measured intensity profile shown in the upper panel reflects the convolution of
the Spot-o-Matic PSF and the pixel response function, the latter of which
includes contributions from lateral charge diffusion as well as capacitive
coupling between neighboring pixels. Capacitive coupling is a deterministic
process by which pixels share charge after photon collection. This is in contrast
to charge diffusion which occurs prior to charge collection \citep{Moore05,MattB06} and
is stochastic.

Once the best focus is established, a small region of pixels (approximately $6
\times 6$ pixels) is step-scanned in two dimensions to measure pixel response.
The spot is scanned repeatedly across the detector in the $x$ direction with
incremental steps of $1\,\mu$m to $2\,\mu$m in the $y$ direction between scans.
The detector response is recorded at each spot position to produce detailed
two-dimensional pixel response profiles.

\begin{figure}[htbp]
\begin{center}
\includegraphics[width=0.95\linewidth]{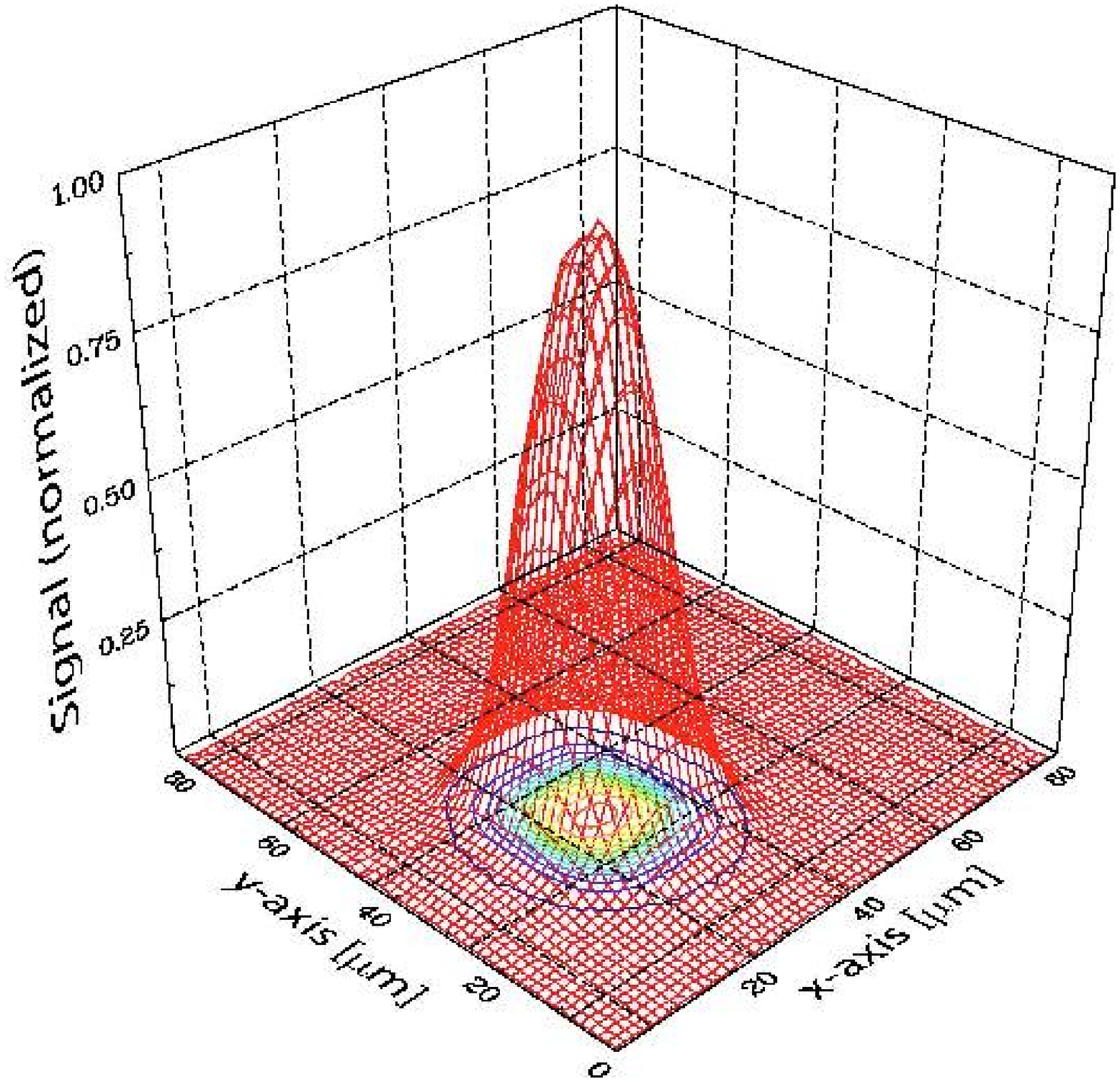}
\caption{Two-dimensional scan of an arbitrary single pixel at a wavelength of
$1050\,$nm.  The grid on the bottom represents the physical size of the pixel.}
\label{Fig:2d-virtual-knife-edge-scans}
\end{center}
\end{figure}

The measured single pixel response shown in
Fig.~\ref{Fig:2d-virtual-knife-edge-scans} is nearly symmetric. Lateral charge
diffusion, capacitive coupling, as well as a contribution from the higher order
rings of the Airy disk produce a signal in the measured pixel even when the spot
is projected onto a neighbor pixel. Diffraction accounts for only a small portion
of the signal measured outside the illuminated pixel: when a spot (with a
wavelength of $1050\,$nm) is centered in a pixel, approximately $2\%$ of the
light is diffracted onto the eight surrounding pixels. Almost all of the  signal
measured outside the illuminated pixel is due to charge diffusion and capacitive
coupling between the pixels. Charge diffusion increases the edge transition's rms
width to $\sigma=2.6\,\mu$m, compared to $\sigma=0.95\,\mu$m obtained from the
Foucault knife-edge scan.

Extraction of the PRF requires unfolding of the measured pixel response from the
measured point spread function of the projected spot (see
Section~\ref{sub:deconv}).

\section{Results and Analysis}
\label{sec:results}

The pixel scan data is analyzed in several ways. First, the data from adjacent
pixels is summed to identify dips in sensitivity between those pixels. The spot
PSF is then removed so that diffusion and capacitive coupling can be measured.
Next, the full two-dimensional scans are summed over an $8 \times 8$ region of
pixels to investigate the integrated response as a function of PSF centroid
position.

\begin{deluxetable}{lllcl}
\tablecolumns{2} \tablewidth{0pc} \tablecaption{Five $1.7\,\mu$m HgCdTe FPAs
have been tested using the Spot-o-Matic.  The top three sensors in the table
have measured quantum efficiencies (QE) over $80\%$, equivalent to nearly
$100\%$ internal quantum efficiency. The bottom two sensors have lower quantum
efficiency and exhibit substantial non-uniform intra-pixel response.}
\tablehead{ \colhead{NIR }    & \colhead{Manu- } & \colhead{Device
} & \colhead{Measured } & \colhead{Substrate } \\
\colhead{sensor }    & \colhead{facturer } & \colhead{ID} & \colhead{ QE} & \colhead{removed } }
\startdata
InGaAs & RVS\tablenotemark{a} & Virgo 1k & $>80\%$ & no \\
HgCdTe & RVS & Virgo 598141 & $>80\%$  & partial \\
HgCdTe & Teledyne\tablenotemark{b} & H2RG \#102 & $>80\%$   & yes \\
\hline
HgCdTe & Teledyne & H2RG \#40 & $50-70\%$ & no \\
HgCdTe & Teledyne & H1RG-BA \#25 & $20-30\%$  & yes
\enddata
\label{Table:device-list} \tablenotetext{a}{Raytheon Vision System, Goleta, CA}
\tablenotetext{b}{Teledyne Scientific \& Imaging (formerly Rockwell Science
Center), Camarillo, CA}
\end{deluxetable}

The Spot-o-Matic was used to test five detectors (see Table
\ref{Table:device-list}). All detectors were produced as part of an ongoing
research and development program \citep{Schubnell-SPIE06} and different
performance characteristics were targeted during processing. The first three
devices, which include detectors made of HgCdTe as well as InGaAs, exhibit good
intra-pixel response. Analysis of the two-dimensional summed response profiles
shows that the integrated response is uniform to better than $2\%$ in each of
these detectors. All three devices have nearly $100\%$ internal quantum
efficiency after correcting for reflections at the detector surface. Sections
\ref{sec:102-1d} and \ref{sec:102-2d} present more detailed results for one
HgCdTe device from the first three high quantum efficiency detectors.

The other two devices have much lower quantum efficiency. One of these devices
(H1RG-BA \#25) shows large random deviations (greater than $10\%$) in uniformity.
The other (H2RG \#40) exhibits a periodic structure in pixel response.
Measurements of the intra-pixel structure in H2RG \#40 are presented in
Section~\ref{sub:40}, as this data is used in Section~\ref{sec:photo} to
demonstrate the effects of abnormal pixel response on undersampled point source
photometry.

\subsection{Intra-Pixel Sensitivity Variations in One Dimension}
\label{sec:102-1d}

\begin{figure}[htbp]
\begin{center}
\includegraphics[height=0.95\linewidth,angle=-90]{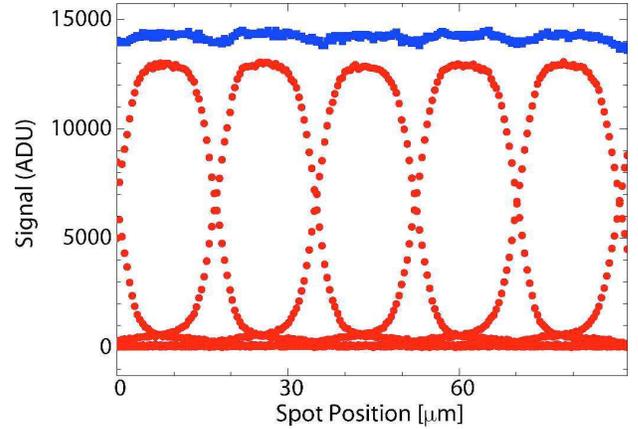}
\caption{One-dimensional scan for $1050\,$nm light over five adjacent pixels
located along the $y$ direction. The scan is performed through the center of
the five pixels. The response of each individual pixel  (filled circles) is
displayed along with the summed response (filled squares) of the five pixels.
The rms fluctuation of the summed response from $20\,\mu$m to $70\,\mu$m is
$1.02\%$.} \label{Fig:3by3-1d-virtual-knife-edge-scans}
\end{center}
\end{figure}

To test for possible loss in sensitivity near pixel boundaries, the responses
of several adjacent pixels are summed, as displayed in
Fig.~\ref{Fig:3by3-1d-virtual-knife-edge-scans}. This integrated response is
then used to estimate total response variation as a function of the PSF
centroid position.

The summed data in Fig.~\ref{Fig:3by3-1d-virtual-knife-edge-scans} has an rms
fluctuation of $1.02\%$. This data shows that, at pixel boundaries, the signal is
shared equally between the two adjacent pixels. This result is typical of all
three high quantum efficiency detectors tested. The data further suggests that
photoelectrons generated at pixel boundaries are collected with close to unit
efficiency. This confirms that lateral charge diffusion or capacitive coupling
\citep{MattB06,Moore05}, rather than inefficient charge collection, is the
dominant source of the intra-pixel variation in this device. In addition, the
tails in the PRF extend far into the neighboring pixel, a clear sign of
capacitive charge sharing.

\subsection{Extracting the Pixel Response}
\label{sub:deconv}

\begin{figure*}[thbp]
\begin{center}
\includegraphics[height=0.45\linewidth, angle=270]{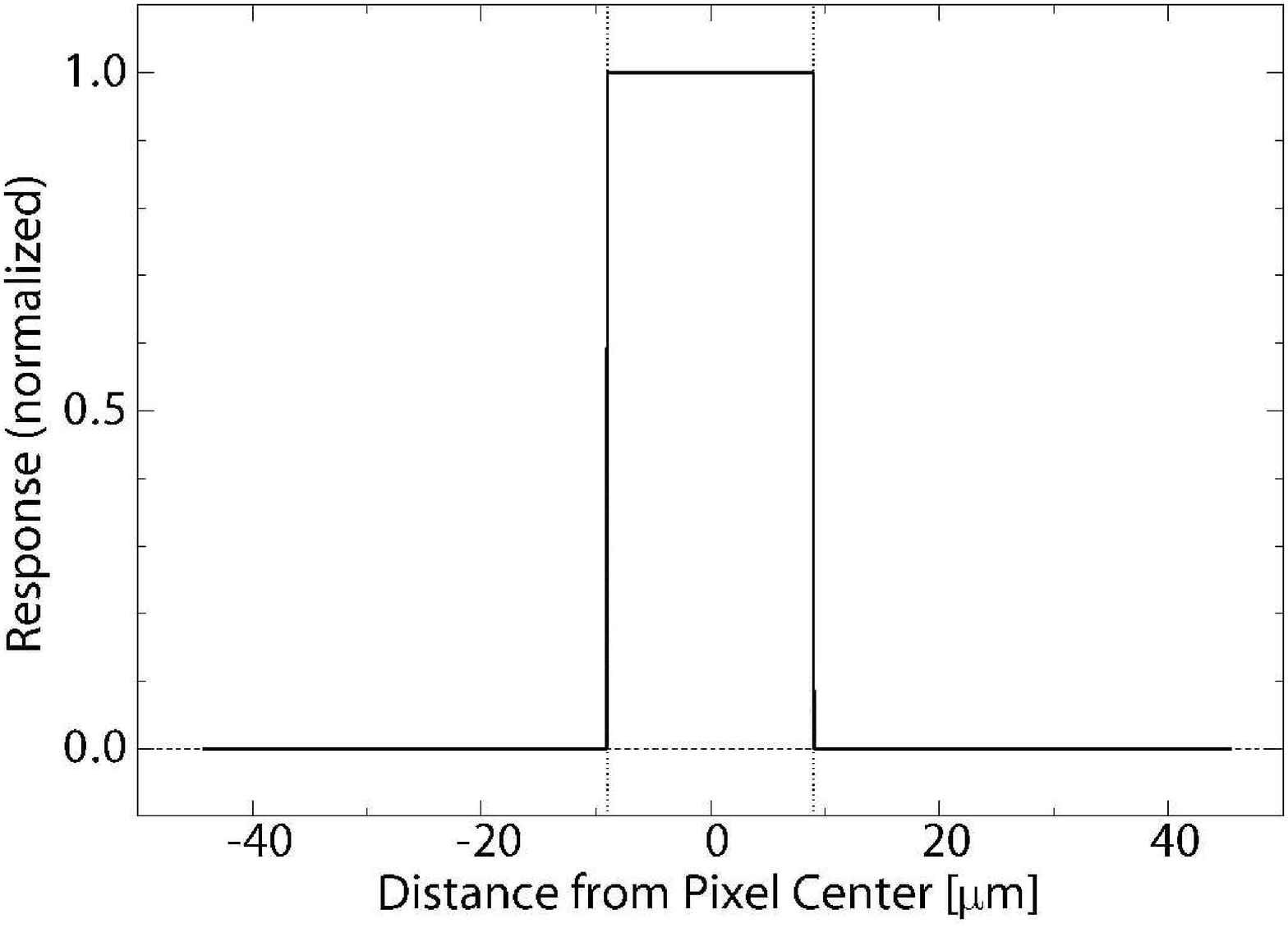}
\ \hspace*{12mm} \
\includegraphics[height=0.45\linewidth, angle=270]{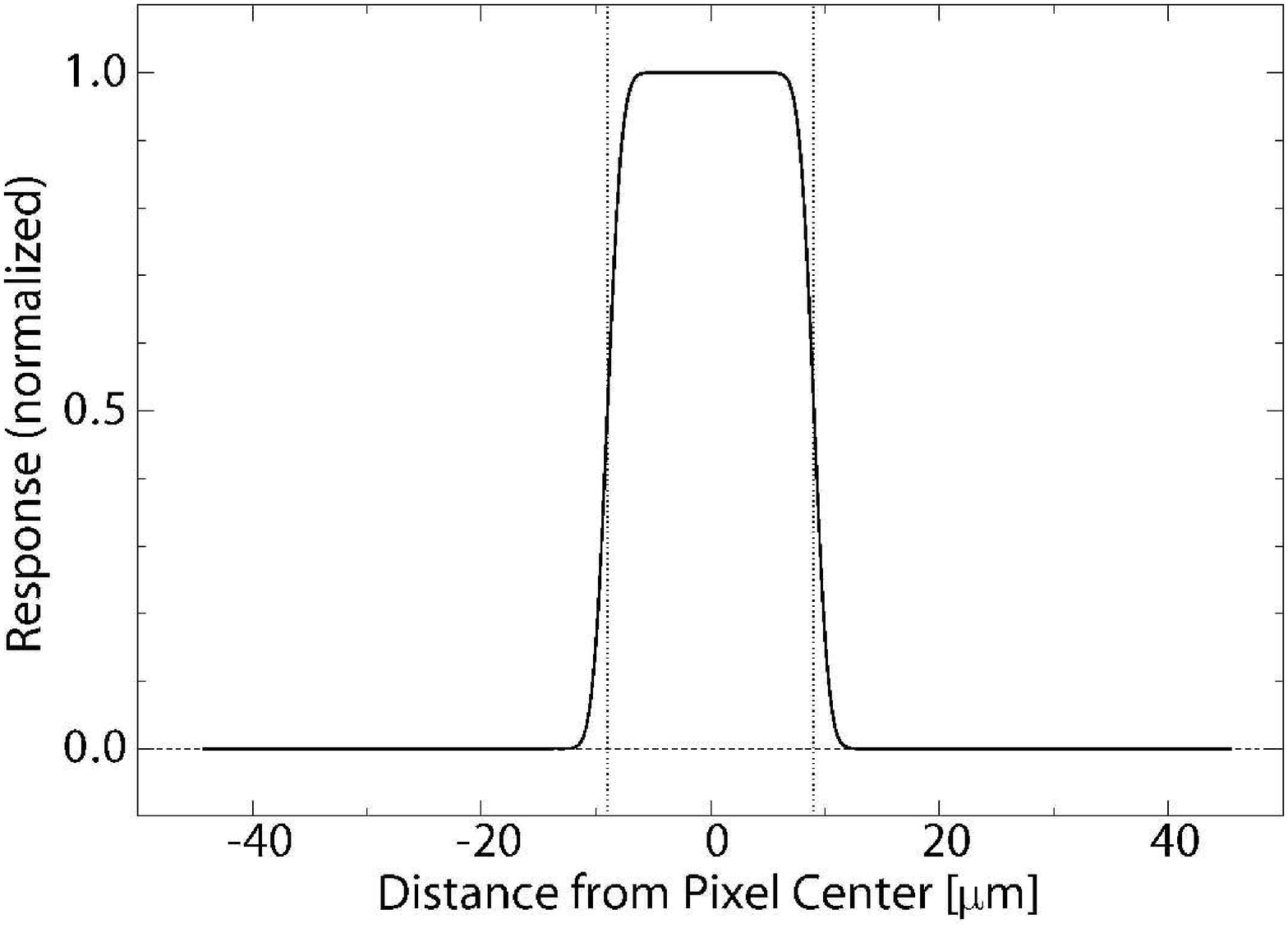}\\[5mm]
\includegraphics[height=0.45\linewidth, angle=270]{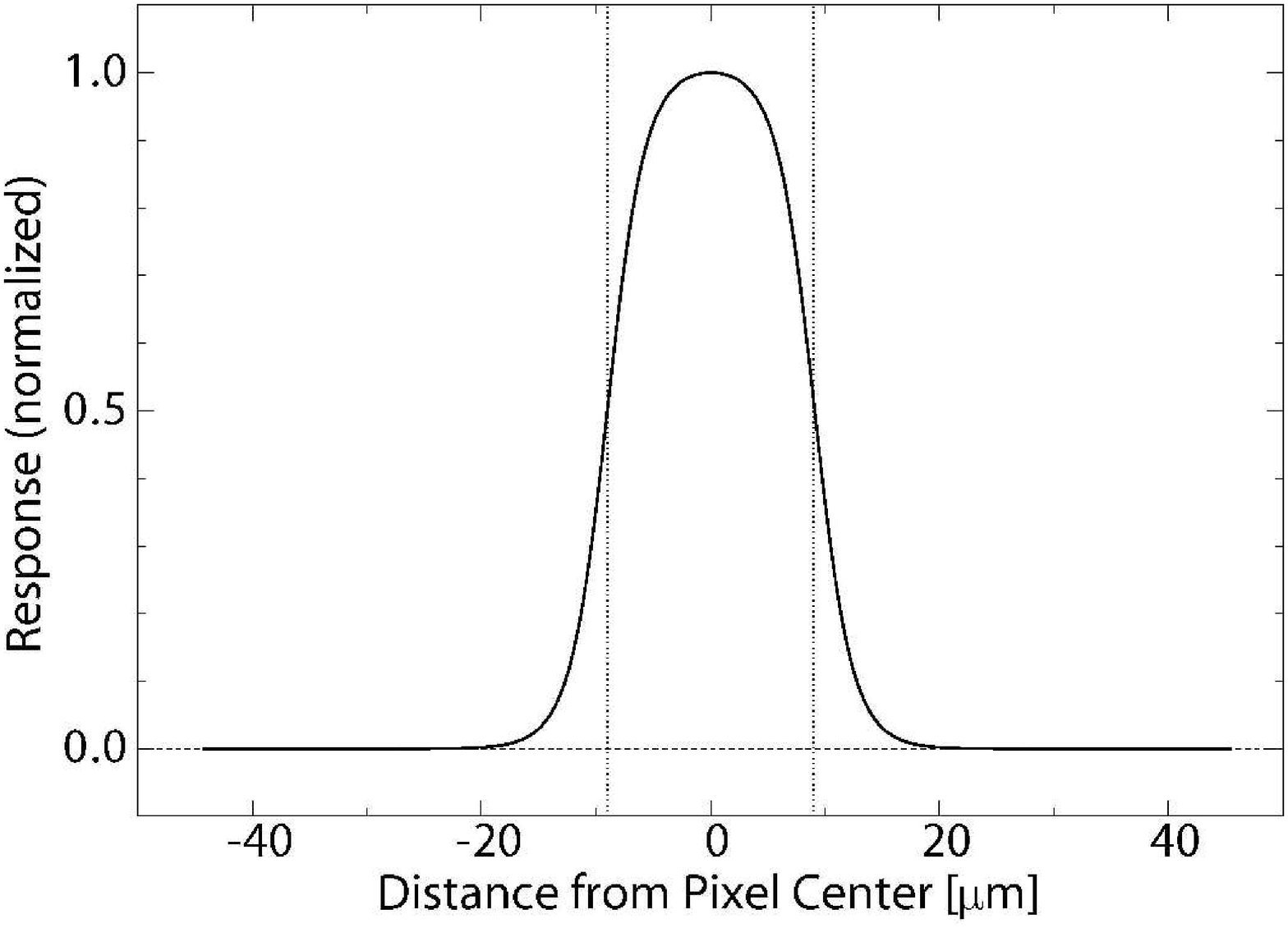}
\ \hspace*{12mm} \
\includegraphics[height=0.45\linewidth, angle=270]{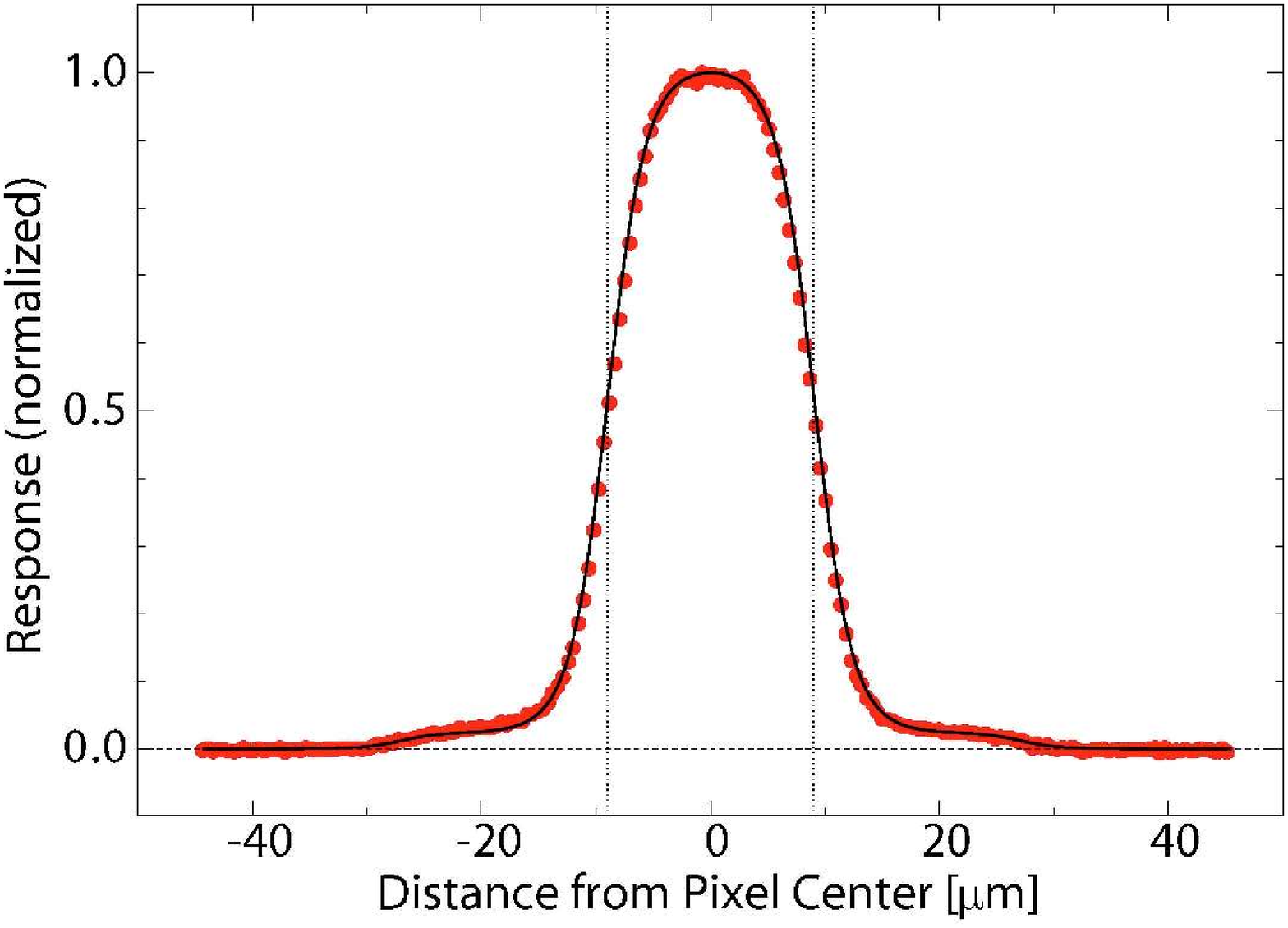}
\caption{Convolution of one-dimensional pixel scan data.  The procedure
begins with
a boxcar PRF (top left), then adds the Spot-o-Matic PSF (top right),
lateral charge diffusion (bottom left) and finally capacitive coupling
(bottom right). The dotted line represents the physical pixel boundary.
The raw data is added in the bottom right panel.}
\label{Fig:convolution}
\end{center}
\end{figure*}

To determine the true pixel response function, the PSF must be unfolded from the
raw data. This allows to understand how charge collection varies within a pixel,
specifically how lateral charge diffusion and capacitive coupling affect the
measured response.

Deconvolution of discretely-sampled data is often difficult due to the small
magnitude of the high-frequency Fourier components. One common method used to
ameliorate this problem is Wiener deconvolution, which adds a small noise term to
each Fourier term.  Wiener deconvolution was attempted to remove the Spot-o-Matic
PSF from the PRF data with limited success. Consequently, an alternative method
was employed which yielded more acceptable results. This method first
approximates each component of the pixel response with a model response function.
It then convolves these components and compares them to the raw data.
Specifically, the detectors are modeled by convolving the measured Spot-o-Matic
PSF with a boxcar response, diffusion, and capacitive coupling, respectively. The
magnitude of the charge diffusion and capacitive coupling are then determined by
fitting this model to the raw data.

The fitting procedure starts with a two parameter (width and position) boxcar
response function\footnote{The width was fixed at the detector pixel pitch
($18\,\mu$m), however, allowing this parameter to vary during the fitting
procedure showed no significant impact on either the best fit value for the
diffusion or the capacitive coupling.}. The boxcar is first convolved with a
Gaussian function with $\sigma = 0.95\, \mu$m, as measured using the Foucault
knife-edge scanning procedure. The result is then convolved with a diffusion term
proportional to the hyperbolic secant, given as
\begin{equation}
I_D(\Delta x) \propto \mbox{sech}(\Delta x/ l_d),
\end{equation}
where $l_d$ is the diffusion length and $\Delta x$ is the distance of the
collected charge from the location of the electron-hole pair. Finally, capacitive
coupling is added to the model by assuming a grid of identical pixels with a
coupling coefficient $\alpha$. In this case, each pixel gains or loses a charge
of $\alpha$ times the difference between the pixel's value and that of each of
its four neighbors. Figure \ref{Fig:convolution} shows the progression of the
model function from the initial boxcar response to the measured pixel response,
using the best fit parameters.

\begin{figure}[thbp]
\hspace*{0mm}
\includegraphics[height=0.94\linewidth, angle=270]{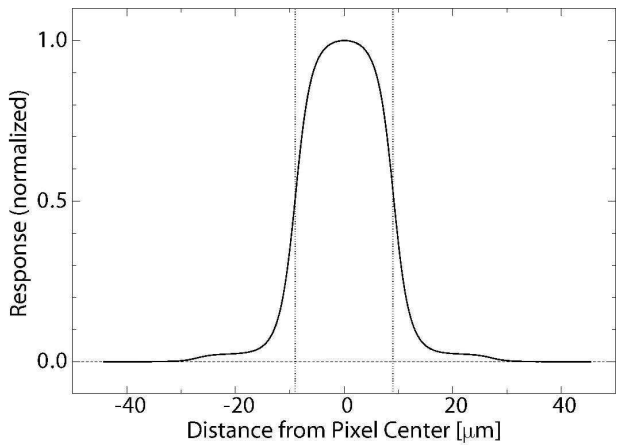}
\caption{A boxcar response convolved with the best fit diffusion and
capacitive coupling components. This is the pixel response function
with the effects of the Spot-o-Matic PSF removed.}
\label{Fig:pixel-fit}
\end{figure}

The extracted pixel response is shown in Fig.~\ref{Fig:pixel-fit}. Note that the
pixel response includes only diffusion and capacitive coupling convolved with a
boxcar response function. The best fit to the raw one-dimensional scan data,
which also includes the Spot-o-Matic PSF, is shown in the bottom right panel of
Fig.~\ref{Fig:convolution}. The pixel responses with and without the spot PSF
included are nearly indistinguishable from each other. The impact of the PSF is
minimal when $\sigma$ is much less than the diffusion length $l_d$. For the pixel
shown, the best fit values for diffusion length and capacitive coupling are $l_d
= 1.87 \pm 0.02\,\mu$m and $\alpha = 2.1 \pm 0.1\,\%$, respectively. From an
independent measurement of the capacitive coupling using the autocorrelation
function, a coupling coefficient of \mbox{$2.2 \pm 0.1\, \%$ \citep{MattB06}} was
obtained, a value which is in excellent agreement with the Spot-o-Matic results.

\subsection{Intra-Pixel Sensitivity Variations in Two Dimensions}
\label{sec:102-2d}

\begin{figure}[htbp]
\begin{center}
\includegraphics[width=0.95\linewidth]{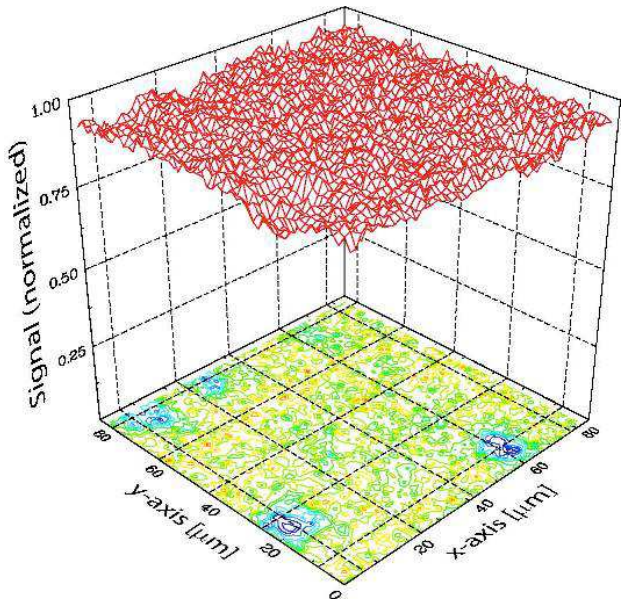}
\caption{Two-dimensional scan at $1050\,$nm
wavelength, summed over an array of $8\times 8$ adjacent
pixels. Only the response of the inner $4\times 4$ array is shown.}
\label{Fig:3by3-2d-virtual-knife-edge-scans}
\end{center}
\end{figure}

The two-dimensional scans from the Spot-o-Matic produce a wealth of information
about both pixel structure and device performance. Figure
\ref{Fig:3by3-2d-virtual-knife-edge-scans} shows a two-dimensional scan extended
over an array of $4 \times 4$ pixels. In order to include all the charge
collected across this array, a $8 \times 8$ array of pixels is summed to produce
this spectrum. The fluctuations in the summed spectrum have an rms deviation of
$1.9\%$. Approximately $1\%$ of the fluctuations in the summed spectrum are
statistical, due to the large thermal background created by the warm optics
radiating through the dewar window. Subtracting this noise in quadrature, the
intra-pixel sensitivity variations are $1.6\%$. When this detector response is
convolved with a critically sampled PSF\footnote{Critical sampling is defined as
a PSF size (FWHM) equal to two resolution elements (e.g. pixels).}, variations of
this magnitude have no measurable effect on precision photometry (see
Section~\ref{sec:photo}).

The noise can be reduced by averaging the response over many exposures at each
position. However, this procedure is not necessary here, as the statistical
uncertainty limit of $1\%$ is sufficient to achieve the goals of the measurement.
The statistical fluctuations quickly average out when convolving the
two-dimensional response functions with larger point spread functions.

Figure~\ref{Fig:3by3-2d-virtual-knife-edge-scans} shows a two-dimensional scan
for a wavelength of $1050\,$nm. The four small dark patches in the contours in
Fig.~\ref{Fig:3by3-2d-virtual-knife-edge-scans} correspond to a drop in
sensitivity of approximately $5\%$. These dips in sensitivity could be due to
small dust particles on the detector surface or defects in the HgCdTe which lead
to traps or recombination.  The same dips in sensitivity are reproduced in scans
using a $1550\,$nm wavelength (not shown). These small dips are not apparent in
flat-field images and do not impact the photometric precision. However, they do
show that the Spot-o-Matic can detect micron-sized variations at the percent
level. They also demonstrate that a simple addition of adjacent pixels restores
photometric precision to better than $2\%$, despite these dips in sensitivity.

\subsection{A sensor with anomalous substructure}
\label{sub:40}

All of the high quantum efficiency detectors have uniform pixel response and
small ($<2\%\;\rm rms$) deviations in the summed spectrum. Detectors with low
quantum efficiency were expected to exhibit large random fluctuations (as in
H1RG-BA \# 25) or dips in sensitivity near pixel edges, as observed by Finger
et al. \citep{Finger}. An unexpected result was discovered in an early
engineering grade device, H2RG \# 40,  which shows an anomalous intra-pixel
structure. This detector displayed close to the best performance for a
$1.7\,\mu$m HgCdTe detector at the time it was fabricated. The read noise (35
electrons using Fowler-1), quantum efficiency ($50-70\%$) and dark current
(0.05 electrons/pixel/second) were typical of the best performance achieved in
developmental FPAs for the \emph{Hubble Space Telescope's} Wide Field Camera 3
upgrade \citep{WFC3}, but none of these tests indicated any potential problems
with the pixel response.

\begin{figure}[htbp]
\begin{center}
\includegraphics[height=0.95\linewidth,angle=-90]{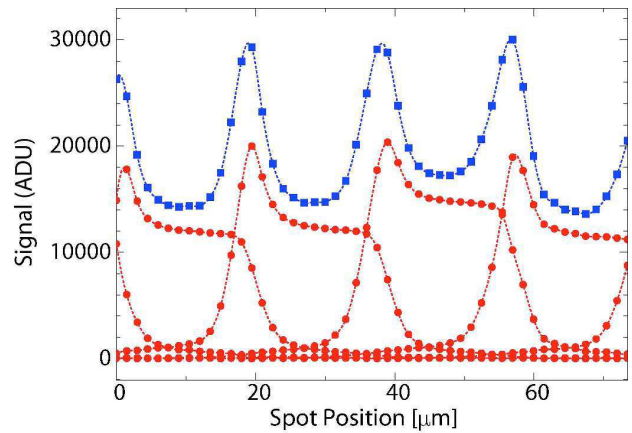}
\caption{One-dimensional scan over four adjacent pixels located along the $y$
direction. The scan is performed through the center of the four pixels. The
response of each individual pixel (filled circles) is displayed, together with
the summed response (filled squares) of the four pixels.}
\label{1-d-chair-rockwell-40}
\end{center}
\end{figure}

\begin{figure}[htbp]
\begin{center}
\bigskip
\includegraphics[width=0.95\linewidth]{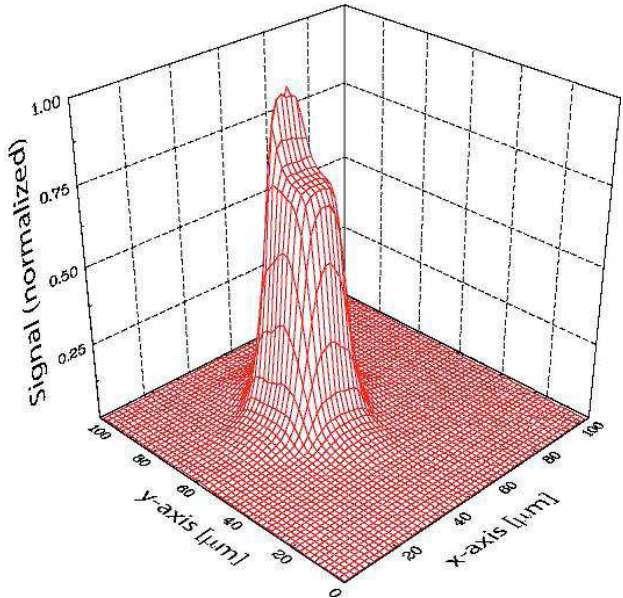}
\caption{Two-dimensional scan of an arbitrary single pixel at a wavelength of
$1300\,$nm for H2RG \#40.} \label{2-d-chair-rockwell-40}
\end{center}
\end{figure}

One-dimensional scans of this device showed an unexpected asymmetric intra-pixel
response (see Fig.~\ref{1-d-chair-rockwell-40}). The two-dimensional profile of
an individual pixel in Fig.~\ref{2-d-chair-rockwell-40} revealed this puzzling
"chair-like" structure in greater detail. This same structure is present in all
the pixels tested on this detector.  Three distinct regions, including two near
the edge and one near the center of the detector, were sampled with the
Spot-o-Matic and all exhibit a similar intra-pixel response.

\begin{figure}[htbp]
\begin{center}
\bigskip
\includegraphics[width=0.95\linewidth]{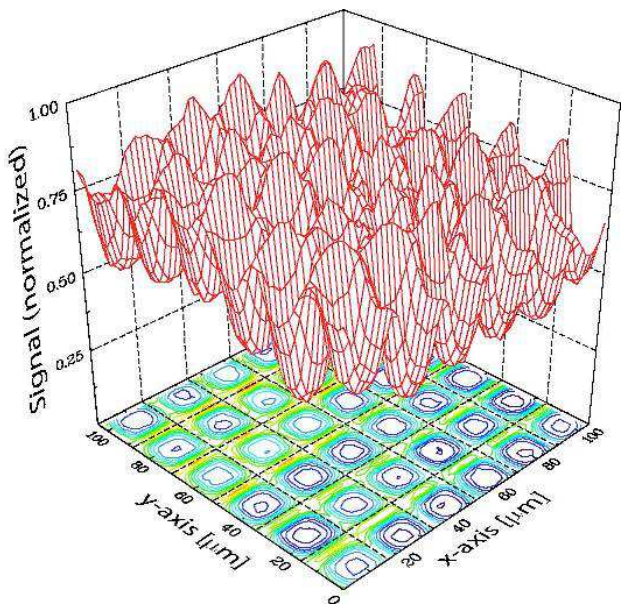}
\caption{Two-dimensional scan summed over a $6\times 6$ array of adjacent pixels
for H2RG \#40.} \label{2-d-chair-rockwell-40-sum}
\end{center}
\end{figure}

The summed response of this device exhibits the periodic peaks and valleys shown
in Fig.~\ref{2-d-chair-rockwell-40-sum}. The rms variation in the summed spectrum
is $18\%$. Since measurements of typical device characteristics average out any
intra-pixel variations, these measurements would not capture the anomalous
substructure revealed by the Spot-o-Matic measurement. Yet such intra-pixel
sensitivity variations can significantly degrade photometry in undersampled
observations, and thus are important to detect.

\section{Photometry Simulations}
\label{sec:photo}

The single pixel response functions and two-dimensional summed scans produced
with the Spot-o-Matic can be used to simulate photometry errors for a range of
different PSF widths. For critically or oversampled point spread functions,
intra-pixel variations have little impact on photometry. However, for
undersampled images, sensitivity variations (such as those observed in FPA H2RG
\#40) can lead to large photometry errors. Assuming modest undersampling (e.g. a
factor of three), the intra-pixel variation in this particular device would
result in rms photometry errors of about 5\%.

\begin{figure}[thbp]
\begin{center}
\bigskip
\includegraphics[height=0.95\linewidth, angle=270]{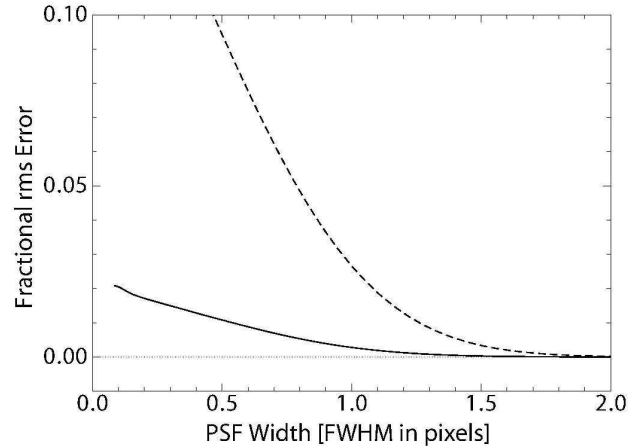}
\caption{Fractional photometric error vs.~PSF size for a typical high quantum
efficiency detector (solid curve), and FPA \#40, with $50\%$ to $70\%$ quantum
efficiency and an anomalous substructure in the pixel response (dashed curve).}
\label{fig:photo-errors}
\end{center}
\end{figure}

Figure~\ref{fig:photo-errors} shows the fractional rms error as a function of
PSF size for the two detectors profiled in Section~\ref{sec:results}.  The
results shown in this figure assume a grid of identical pixels with the
two-dimensional response profiles shown in
Figs.~\ref{Fig:2d-virtual-knife-edge-scans} and \ref{2-d-chair-rockwell-40}.
The summed response spectrum of the pixel grid is convolved with a Gaussian
point spread function with FWHM values ranging from a fraction of a pixel to
two pixels.  For the three high quantum efficiency detectors tested, the
photometric errors are less than $2\%$ for any size PSF. However, if a detector
has a substructure as displayed in Fig.~\ref{2-d-chair-rockwell-40-sum}, the
photometric errors may be well over $10\%$ when the PSF size is much less than
one pixel. As the PSF size increases, the intra-pixel variations average out,
and for a PSF size of more than two pixels the photometric errors are
negligible.

\section{Summary}

The automated point projection system described here provides the ability to
detect and accurately characterize substructure in the pixel response of focal
plane detectors. While the measurements were limited to NIR detector arrays, the
Spot-o-Matic can also be used for the measurement of sub-pixel structure in CCDs.

The Spot-o-Matic has been tested with five detectors from both Raytheon Vision
Systems and Teledyne Scientific \& Imaging.  The results for devices with near
$100\%$ internal quantum efficiency indicate that the pixel response is uniform
to better than $2\%$ in all areas tested. This result is not surprising; high
quantum efficiency detectors must count nearly all of the incident photons.

By contrast, a detector with moderate quantum efficiency, and reasonable read
noise and dark current levels has exhibited a strong asymmetric intra-pixel
structure. This otherwise high quality detector would cause large photometric
errors in an undersampled instrument. Spot-o-Matic measurements can discover and
characterize variations of this nature. For undersampled imaging such a detailed
understanding of the intra-pixel structure is essential for obtaining precise
photometric measurements.

\acknowledgments

This work was supported by DOE grant No. DE-FG02-95ER40899.

\end{document}